\documentclass[final,5p,times,twocolumn,numbers]{elsarticle}

\setcitestyle{square}

\usepackage{graphicx}
\usepackage{mathtools,amssymb} 
\usepackage{physics}
\usepackage{dcolumn} 
\usepackage{bm} 
\usepackage{xcolor}
\usepackage{dsfont} 
\usepackage{slashed} 
\usepackage[hidelinks]{hyperref}

\newcommand \beq{\begin{eqnarray}}
\newcommand \eeq{\end{eqnarray}}

\journal{Physics Letters B}

\begin{document}

\begin{frontmatter}

\title{Mesons, baryons and the confinement/deconfinement transition}

\author[CPhT]{V. Tomas Mari Surkau\corref{c1}}
\ead{victor-tomas.mari-surkau@polytechnique.edu}

\author[CPhT]{Urko Reinosa\corref{c2}}
\ead{urko.reinosa@polytechnique.edu}

\cortext[c1]{Corresponding author}

\affiliation[CPhT]{organization={Centre de Physique Th\unexpanded{\'e}orique, CNRS, \unexpanded{\'E}cole Polytechnique, IP Paris, F-91128 Palaiseau, France.}}

\begin{abstract}
We identify an observable that could operate as a probe of the quark versus hadron content of a bath of quarks and gluons at finite temperature and chemical potential. To this purpose, we relate the Polyakov and anti-Polyakov loops, which determine how energetically costly it is to bring an external static quark or antiquark probe into the thermal bath, to the ability of that medium to provide favorable conditions for the formation of meson-like or baryon-like configurations that would screen the probes.
\end{abstract}

\begin{keyword}
Quantum chromodynamics \sep Confinement \sep QCD phase diagram \sep Net quark number

\end{keyword}

\end{frontmatter}

Since the advent of Quantum Chromodynamics (QCD) as the theory for the strong interaction, one mystery prevails: \textit{How do its elementary degrees of freedom, the quarks and the gluons, turn into the actual mesons and baryons observed in experiments?}
One popular way to attack the problem is to consider QCD as a thermodynamic system and to study its properties as functions of external parameters such as temperature $T$ and/or density. In this context, lattice simulations strongly suggest that the low- and high-temperature regimes of QCD are controlled by distinct active degrees of freedom, hadrons on the one side, and quarks and gluons on the other side \cite{Aarts2023PhasePhysics}. These two regimes are not separated by a sharp transition, however, but rather by a crossover \cite{Aoki2006ThePhysics}.

Still, in the formal limit where all quarks are considered infinitely heavy, this crossover turns into an actual phase transition, associated with the breaking of a symmetry, known as the center symmetry \cite{Svetitsky1982CriticalTransitions}, and probed by an order parameter, the Polyakov loop $\ell$ \cite{Polyakov1978ThermalLiberation}. The latter gives access to the energy cost ${\Delta F\equiv -T\ln\ell}$ of bringing a static quark probe into a thermal bath of gluons. A vanishing Polyakov loop, associated with an explicitly realized center symmetry, corresponds to the confined phase of the system, where the addition of an isolated quark is forbidden. For finite but large quark masses, the thermal bath contains both quarks and gluons. Still, the Polyakov loop keeps its interpretation and remains a good probe of the distinct phases, as it is small in the low-temperature phase, indicating that adding a static quark probe, though not forbidden anymore, remains energetically very costly.

A natural question arises though: How can the possibility of adding a quark into the low-temperature bath be compatible with the change in the relevant degrees of freedom mentioned above? In this Letter, we argue that the possibility of adding an external static quark probe to the low-temperature bath is deeply connected to the tendency of that same medium to form mesons or baryons. To this purpose, we determine the net quark number gained by the system in the presence of a quark or antiquark probe compared to that of the system in the absence of the probe.

We argue that, in the high-temperature, deconfined phase, the net quark number gained by the system is asymptotically equal to that of the color probe, in line with the fact that the relevant degrees of freedom in this case are quarks and that the color probe can be added without significantly affecting its average net quark number. In contrast, in the low-temperature, confined phase, we find that, depending on the value of the chemical potential and/or on the considered color probe, the net quark number gained by the system takes the integer values $0$ or $\pm3$ in line with the fact that the relevant degrees of freedom are not quarks anymore and that the medium rearranges itself to incorporate the color charge within acceptable, meson-like or baryon-like states. The terminology emphasizes the fact that our analysis does not give access to the color representation of the formed states, nor does it inform us of their localization in space. Thus, in a sense, our result checks one of the necessary conditions for the formation of actual mesons or baryons.\footnote{See below for a possible refinement of the analysis.}

We work in the heavy-quark regime where the Polyakov and anti-Polyakov loops are the most relevant since they allow for a sharp distinction between the two phases. Our argumentation is largely model independent as it relies only on the one-loop expression for the matter contribution to the Polyakov loop thermodynamic potential, a good approximation in the heavy-quark regime, and on the well-established fact that the gluonic contribution is confining \cite{McLerran:1981pb, Brown:1988qe, Gupta:2007ax, Braun:2007bx}. We also argue that some of our findings should hold in the full QCD case.

\section{The Polyakov loop potential}\label{Xsec1}
Let us then consider a bath of quarks and gluons at temperature $T$ and quark chemical potential $\mu$, and study the net quark number gained by the system upon bringing a static quark or antiquark probe, defined as the quark number of the probe plus the net quark number response of the bath in the presence of the probe. The latter can be related to the Polyakov loops as follows.\footnote{ We stress that our use of the expression ``Polyakov loop'' is slightly abusive (although quite common in the literature) since what we actually consider is the thermal average ${\ell(\vec{x})\equiv\langle\Phi[A_0](\vec{x})\rangle}$ of the Polyakov loop defined as the trace of a time-ordered exponential along the compact temporal direction:
\begin{equation*}\Phi[A_0](\vec{x})\equiv{\rm tr}\,{\cal P}\,\exp\left\{i\int_0^\beta d\tau\,A_0^a(\tau,\vec{x})\,t^a\right\}.\end{equation*}} We know that the increase $\Delta F_q$ (resp. $\Delta F_{\bar q}$) in the free-energy\footnote{For a system at finite chemical potential, we have in mind the Landau free-energy, also known as grand potential. It is usually denoted $\Omega$ but here we keep the notation $F$ for simplicity.} of the bath upon bringing the quark probe (resp. the antiquark probe) is related to the Polyakov loop $\ell$ (resp. the anti-Polyakov loop $\bar\ell$) as
\beq
\Delta F_q=-T\ln \ell, \quad {\rm resp.} \quad \Delta F_{\bar q}=-T\ln \bar\ell\,.\label{eq:DF}
\eeq
Then, using the well known relation ${Q=-\partial F/\partial\mu}$ between the free-energy of the system and the average value of a conserved charge $Q$ of associated chemical potential $\mu$, we deduce from Eq.~(\ref{eq:DF}) that the net quark number response $\Delta Q_q$ (resp. $\Delta Q_{\bar q}$) of the bath is
\beq
\Delta Q_q=\frac{T}{\ell}\frac{\partial\ell}{\partial\mu}, \quad {\rm resp.} \quad \Delta Q_{\bar q}=\frac{T}{\bar\ell}\frac{\partial\bar\ell}{\partial\mu},\label{eq:Q}
\eeq
and thus that the net quark number gained by the system (including the probe) upon bringing a quark probe is ${\Delta Q_q+1}$, while the net quark number gained by the system (including the probe) upon bringing an antiquark is ${\Delta Q_{\bar q}-1}$. Note that, because the Polyakov loops are gauge-invariant and renormalize multiplicatively, the net quark number gains are both gauge-invariant and renormalization-group-invariant, which qualifies them as good theoretical observables. The question of whether these could correspond to actual experimental observables is, of course, a more subtle one, because it is far from obvious what it means, experimentally, to bring an isolated color probe into the medium.

The Polyakov loops needed in Eq.~\eqref{eq:Q} are obtained from the extremization\footnote{For non-zero, real chemical potential, the relevant extremum is usually a saddle point in the plane ${(\ell,\bar\ell)\in\mathds{R}^2}$, with ${\ell\neq\bar\ell}$. In what follows, it will be crucial not to consider the simplification ${\bar\ell=\ell}$ sometimes used in the literature.}  of the Polyakov loop effective action $\Gamma[\ell,\bar\ell]$, where $\ell$ and $\bar\ell$ denote position dependent fields a priori. Under the assumption of unbroken translation invariance,\footnote{It would of course be interesting to extend the present analysis to phases with broken translation invariance.} the values of these fields at the extremum of $\Gamma[\ell,\bar\ell]$ are position independent. This means that, in practice, it is enough to extremize the Polyakov loop potential $V(\ell,\bar\ell)$, obtained from $\Gamma[\ell,\bar\ell]$ by restricting to homogeneous fields. In principle, $V(\ell,\bar\ell)$ can be systematically computed order by order using a saddle-point expansion around a constant gluonic background; see Refs.~\cite{Dumitru:2013xna,Guo:2018scp}. See the Appendix for alternative approaches that also use constant gluonic backgrounds.

Without loss of generality, $V(\ell,\bar\ell)$ can be written as
\beq
V(\ell,\bar\ell)=V_{\rm glue}(\ell,\bar\ell)+V_{\rm quark}(\ell,\bar\ell)\,.\label{eq:V}
\eeq
The glue part $V_{\rm glue}(\ell,\bar\ell)$ stands for the sum of contributions to $V(\ell,\bar\ell)$ not involving quarks and is then nothing but the Polyakov loop potential in the associated pure gauge theory. As for the quark part $V_{\rm quark}(\ell,\bar\ell)$, it is made of all the contributions to $V(\ell,\bar\ell)$ that do involve quarks. Let us now delimitate the set-up and approximations considered for  $V_{\rm glue}(\ell,\bar\ell)$ and $V_{\rm quark}(\ell,\bar\ell)$ in this work.

As far as $V_{\rm glue}(\ell,\bar\ell)$ is concerned, we do not use any particular approximation, but, instead, rely on a few, well-established properties of the associated pure gauge theory. In particular, $V_{\rm glue}(\ell,\bar\ell)$ is center-symmetric,
\beq
V_{\rm glue}(\ell,\bar\ell)=V_{\rm glue}(e^{i2\pi/3}\ell,e^{-i2\pi/3}\bar\ell),\label{eq: center}
\eeq
and confining at low temperatures. By confining, we mean that the relevant extremum of $V_{\rm glue}(\ell,\bar\ell)$ is located at the center-symmetric or confining point ${(\ell,\bar\ell)=(0,0)}$ in this limit.

As for $V_{\rm quark}(\ell,\bar\ell)$, for simplicity, we consider $N_f$ degenerate quark flavors of mass $M$, but the discussion can easily be extended to nondegenerate flavors. We also consider the heavy-quark regime where this mass $M$ is taken large. In this range, we approximate $V_{\rm quark}(\ell,\bar\ell)$ by the one-loop expression \cite{Fukushima2004ChiralLoop}
\beq\label{eq: quark}
& & V_{\rm quark}(\ell,\bar\ell)=-\frac{TN_f}{\pi^2}\int_0^\infty dq\,q^2\nonumber\\
& & \hspace{0.2cm}\times\,\Bigg\{\ln\Big[1\!+\!3\ell e^{-\beta(\varepsilon_q-\mu)}\!+\!3\bar\ell e^{-2\beta(\varepsilon_q-\mu)}\!+\!e^{-3\beta(\varepsilon_q-\mu)}\Big]\nonumber\\
& & \hspace{0.5cm}+\ln\Big[1\!+\!3\bar\ell e^{-\beta(\varepsilon_q+\mu)}\!+\!3\ell e^{-2\beta(\varepsilon_q+\mu)}\!+\!e^{-3\beta(\varepsilon_q+\mu)}\Big]\,\Bigg\}\,,
\eeq
with ${\beta\equiv 1/T}$ and ${\varepsilon_q\equiv\sqrt{q^2+M^2}}$. This is a rather common approximation in the heavy-quark regime, which captures a number of features of the phase structure \cite{Kashiwa:2012wa,Reinosa:2015oua,Maelger:2018vow}. In the context of Landau-gauge-fixed calculations, it has even been argued that this is a controlled approximation once the non-perturbative gauge-fixing issues of the glue sector have been properly taken into account or modeled \cite{Barrios:2021cks}. Let us also stress that the one-loop approximation is certainly controlled at high temperatures owing to asymptotic freedom. The study of the role of higher order corrections at low temperatures is beyond the scope of the present work. It would be interesting in particular to investigate whether the two-loop corrections \cite{Maelger:2017amh} in this limit are dominated by linear combination of $\ell e^{-\beta(M-\mu)}$, $\bar\ell e^{-2\beta(M-\mu)}$, $\bar\ell e^{-\beta(M+\mu)}$ and $\ell e^{-2\beta(M+\mu)}$ in which case the main result of this work, see below, should not be modified. We will report on this analysis elsewhere.

We finally need to make some assumption on the relation between $V_{\rm glue}(\ell,\bar\ell)$ and $V_{\rm quark}(\ell,\bar\ell)$. We assume that, in the zero-temperature limit, and as long as ${|\mu|<M}$, the quark contribution is suppressed with respect to the glue contribution. This is due to the fact that the former is exponentially suppressed, as shown below, and that, according to various continuum studies \cite{Braun:2007bx, Reinosa:2014ooa}, confinement is triggered by the presence of massless modes in the glue potential, which makes it behave as a power law at low temperatures. Such power-law behaviors are also a common feature of popular models for the glue potential, such as the ones in Refs.~\cite{Reinosa:2014ooa,Ratti:2005jh, Lo:2013hla, MariavanEgmond2022ATemperature, MariSurkau2024DeconfinementDependences}. The situation differs for ${|\mu|\geq M}$, as we discuss below.

Let us now argue that the behavior of the net quark number gained by the system at low and high temperatures can be inferred from the few ingredients we have just enumerated.

\section{Low temperatures}\label{Xsec2}

The equations determining $\ell$ and $\bar\ell$ are ${\partial V/\partial\ell=0}$ and ${\partial V/\partial\bar\ell=0}$. Using Eqs.~(\ref{eq:V}) and (\ref{eq: quark}), they write
\begin{eqnarray}
\frac{\partial V_{\rm glue}}{\partial\ell} & \!\!\!\!=\!\!\!\! & \frac{TN_f}{\pi^2}\int_0^\infty dq\,q^2\nonumber\\
& \!\!\!\!\times\!\!\!\! & \Bigg\{\frac{3 e^{-\beta(\varepsilon_q-\mu)}}{1\!+\!3\ell e^{-\beta(\varepsilon_q-\mu)}\!+\!3\bar\ell e^{-2\beta(\varepsilon_q-\mu)}\!+\!e^{-3\beta(\varepsilon_q-\mu)}}\nonumber\\
& & \hspace{0.1cm}+\,\frac{3 e^{-2\beta(\varepsilon_q+\mu)}}{1\!+\!3\bar\ell e^{-\beta(\varepsilon_q+\mu)}\!+\!3\ell e^{-2\beta(\varepsilon_q+\mu)}\!+e^{-3\beta(\varepsilon_q+\mu)}}\Bigg\}\,,\label{eq:g1}
\end{eqnarray}
and
\begin{eqnarray}
\frac{\partial V_{\rm glue}}{\partial\bar\ell} & \!\!\!\!=\!\!\!\! & \frac{TN_f}{\pi^2}\int_0^\infty dq\,q^2\nonumber\\
& \!\!\!\!\times\!\!\!\! & \Bigg\{\frac{3 e^{-2\beta(\varepsilon_q-\mu)}}{1\!+\!3\ell e^{-\beta(\varepsilon_q-\mu)}\!+\!3\bar\ell e^{-2\beta(\varepsilon_q-\mu)}\!+\!e^{-3\beta(\varepsilon_q-\mu)}}\nonumber\\
& & \hspace{0.1cm}+\,\frac{3 e^{-\beta(\varepsilon_q+\mu)}}{1\!+\!3\bar\ell e^{-\beta(\varepsilon_q+\mu)}\!+\!3\ell e^{-2\beta(\varepsilon_q+\mu)}\!+\!e^{-3\beta(\varepsilon_q+\mu)}}\Bigg\}\,.\label{eq:g2}
\end{eqnarray}
To deduce the behavior of $\ell$ and $\bar\ell$ as ${T\to 0}$, we need to discuss these equations separately for ${|\mu|<M}$, ${|\mu|> M}$ and ${|\mu|=M}$.

\subsection{$|\mu|<M$}\label{Xsec3}
Any exponential $e^{-n\beta(\varepsilon_q\pm\mu)}$ appearing in the right-hand sides of Eqs.~(\ref{eq:g1}) and (\ref{eq:g2}), with ${n=1}$, $2$ or $3$, is bounded by
\beq
e^{-n\beta(\varepsilon_q\pm\mu)}\leq e^{-n\beta(M-|\mu|)},
\eeq
a bound that is independent of the integration variable $q$ and which goes to $0$ exponentially as ${T\to 0}$ if we assume that ${|\mu|<M}$. This means that, in this case, we can safely approximate the above equations as
\beq
\frac{\partial V_{\rm glue}}{\partial\ell} & \!\!\!\!=\!\!\!\! & \frac{3TN_f}{\pi^2}\int_0^\infty dq\,q^2\Big(e^{-\beta(\varepsilon_q-\mu)}+ e^{-2\beta(\varepsilon_q+\mu)}\Big),\label{eq:g3}\\
\frac{\partial V_{\rm glue}}{\partial\bar\ell} & \!\!\!\!=\!\!\!\! & \frac{3TN_f}{\pi^2}\int_0^\infty dq\,q^2\Big(e^{-\beta(\varepsilon_q+\mu)}+e^{-2\beta(\varepsilon_q-\mu)}\Big),\label{eq:g4}
\eeq
where we note that the quark contribution does not depend on $\ell$ or $\bar\ell$ anymore. Moreover, the $\mu$-dependence can be explicitly pulled out of the integrals. More precisely, upon introducing the function
\beq
f_y\equiv \frac{1}{\pi^2}\int_0^\infty \!\!\! dx\,x^2e^{-y\sqrt{x^2+1}}\sim \frac{y^{-3/2}}{\sqrt{2}\pi^{3/2}}e^{-y},\label{eq:f}
\eeq
where the asymptotic expression on the right-hand side is understood in the limit ${y\to\infty}$, the equations fixing $\ell$ and $\bar\ell$ become
\beq
\frac{\partial V_{\rm glue}}{\partial\ell} & \!\!\!\!\simeq\!\!\!\! & C\,\big(e^{\beta\mu} f_{\beta M}+e^{-2\beta\mu} f_{2\beta M}\big),\label{eq:gap1}\\
\frac{\partial V_{\rm glue}}{\partial\bar\ell} & \!\!\!\!\simeq\!\!\!\! & C\,\big(e^{-\beta\mu} f_{\beta M}+e^{2\beta\mu} f_{2\beta M}\big),\label{eq:gap2}
\eeq
with ${C\equiv 3N_f T M^3}$.

For ${|\mu|<M}$, each right-hand side approaches $0$ exponentially as ${T\to 0}$, and, because we have assumed that the glue potential dominates over the exponentially suppressed quark contribution in this limit and is confining, we deduce that $(\ell,\bar\ell)$ approaches $(0,0)$. The left-hand sides of Eqs.~\eqref{eq:gap1} and \eqref{eq:gap2} can then be linearized around ${(\ell,\bar\ell)=(0,0)}$. We find
\beq
& &\left(
\begin{array}{@{}ll@{}}
\partial^2_\ell V_{\rm glue} & \partial_\ell\partial_{\bar\ell} V_{\rm glue}\\
\partial_{\bar\ell}\partial_\ell V_{\rm glue} & \partial^2_{\bar\ell} V_{\rm glue}
\end{array}\right)
\left(
\begin{array}{@{}l@{}}
\ell\\
\bar\ell
\end{array}
\right)\nonumber\\
& & \hspace{1.0cm}\simeq\,
C\left(
\begin{array}{@{}l@{}}
e^{\beta\mu} f_{\beta M}+e^{-2\beta\mu} f_{2\beta M}\\
e^{-\beta\mu} f_{\beta M}+e^{2\beta\mu} f_{2\beta M}
    \end{array}
\right),\label{eq:gap3}
\eeq
where the matrix of second derivatives of $V_{\rm glue}$ is to be taken at ${(\ell,\bar\ell)=(0,0)}$ and we used that the first derivatives vanish at this point due to the center symmetry \eqref{eq: center}. As discussed below, the linearization in Eq.~\eqref{eq:gap3} is not fully consistent. Still, it leads to a good qualitative picture of what happens at small temperatures, which becomes quantitatively accurate in the limit ${T\to 0}$. The outcome of a more consistent analysis will be presented below, but, for now, we shall stick to the linearized approximation as the presentation is simpler.

The symmetry \eqref{eq: center} also implies that both $\partial^2_\ell V_{\rm glue}$ and $\partial^2_{\bar\ell} V_{\rm glue}$ vanish at the center-symmetric point and only ${\partial_{\bar\ell}\partial_\ell V_{\rm glue}=\partial_\ell\partial_{\bar\ell} V_{\rm glue}}$ contributes. The matrix in the left-hand side of Eq.~\eqref{eq:gap3} is then easily inverted, yielding
\beq
\left(
\begin{array}{@{}l@{}}
\ell\\
\bar\ell
\end{array}
\right)\simeq\frac{C}{\partial_\ell\partial_{\bar\ell} V_{\rm glue}}\left(
\begin{array}{@{}l@{}}
e^{-\beta\mu} f_{\beta M}+e^{2\beta\mu} f_{2\beta M}\\
e^{\beta\mu} f_{\beta M}+e^{-2\beta\mu} f_{2\beta M}
\end{array}
\right).\label{eq:l}
\eeq
Using Eq.~\eqref{eq:Q}, one then deduces that
\beq
\Delta Q_q & \!\!\!\!\simeq\!\!\!\! & \frac{-e^{-\beta\mu} f_{\beta M}+2e^{2\beta\mu} f_{2\beta M}}{e^{-\beta\mu} f_{\beta M}+e^{2\beta\mu} f_{2\beta M}},\\
\Delta Q_{\bar q} & \!\!\!\!\simeq\!\!\!\! & \frac{e^{\beta\mu} f_{\beta M}-2e^{-2\beta\mu} f_{2\beta M}}{e^{\beta\mu} f_{\beta M}+e^{-2\beta\mu} f_{2\beta M}}\,.
\eeq
These formulas simplify upon evaluating the net quark numbers gained by the system:
\beq
\Delta Q_q+1 & \!\!\!\!\simeq\!\!\!\! & \frac{3}{1+e^{-3\beta\mu} f_{\beta M}/f_{2\beta M}},\label{eq:nice}\\
\Delta Q_{\bar q}-1 & \!\!\!\!\simeq\!\!\!\! & \frac{-3}{1+e^{3\beta\mu} f_{\beta M}/f_{2\beta M}}\,.\label{eq:nice2}
\eeq
As expected from charge conjugation, the expressions for $\Delta Q_q+1$ and $-(\Delta Q_{\bar q}-1)$ can be obtained from one another using ${\mu\to -\mu}$. This follows from the identity ${V(\ell,\bar\ell;\mu)=V(\bar\ell,\ell;-\mu)}$ which derives itself from charge conjugation. Without loss
of generality, one can then concentrate on ${\mu\geq 0}$ and deduce the case ${\mu<0}$ upon applying the appropriate transformations. Alternatively, one can concentrate on $\Delta Q_q+1$ for any value of $\mu$ and deduce $\Delta Q_{\bar q}-1$.

\begin{figure}
    \centering
    \includegraphics[width=0.9\linewidth]{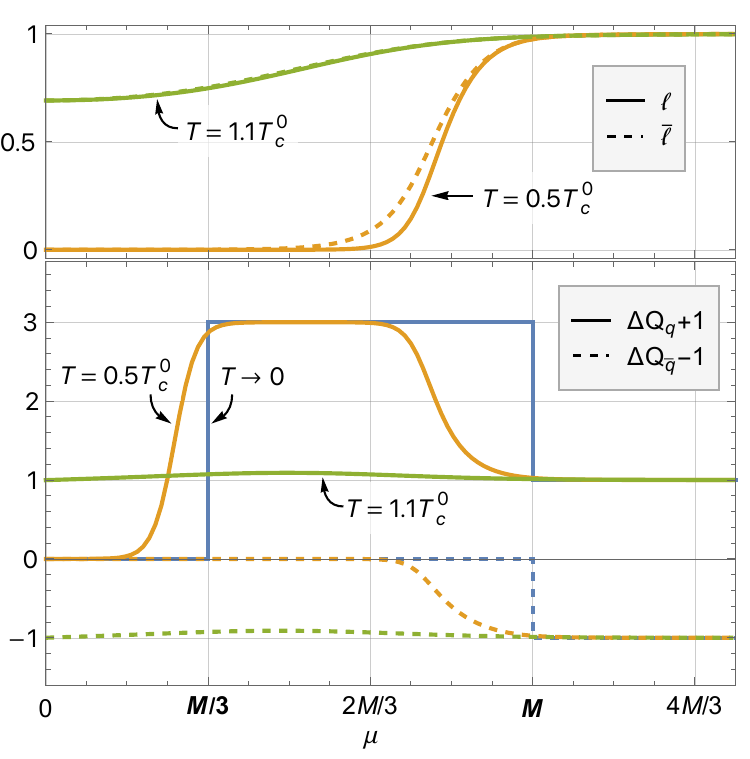}
   \caption{The system's net quark number gains $\smash{\Delta Q_q+1}$ and $\Delta Q_{\bar q}-1$ in the presence of a quark or antiquark probe, as functions of the chemical potential $\mu$ (bottom part), compared to the Polyakov and anti-Polyakov loops $\ell$ and $\bar\ell$ (top part). We considered $\smash{N_f=3}$ degenerate heavy flavors and various temperatures below and above the confinement-deconfinement transition temperature $T_c^0$ at $\smash{\mu=0}$. The corresponding plots for negative $\mu$ can be obtained using the formulas $\smash{\ell(-\mu)=\bar\ell(\mu)}$ and $\Delta Q_q(-\mu)+1=-(\Delta Q_{\bar q}(\mu)-1)$ which are consequences of charge conjugation.}
    \label{fig: Qqbar of mu}
\end{figure}

Recall that the above formulas are qualitative estimates which become exact in the ${T\to0}$ limit. In this limit, using Eq.~\eqref{eq:f}, we see that ${\Delta Q_{\bar q}-1}$ becomes a step function equal to $-3$ for ${\mu<-M/3}$ and $0$ otherwise, while ${\Delta Q_q+1}$ becomes a step function, equal to $0$ for ${\mu<M/3}$ and $3$ otherwise, see Fig.~\ref{fig: Qqbar of mu} for a representation of this limiting step function in the range ${\mu\geq 0}$. We stress that this asymptotic result should hold as long as the glue potential obeys the few basic properties listed above. In particular, Fig.~\ref{fig: Qqbar of mu} also shows the quark number gains for ${\mu\geq 0}$ as obtained using the model of Ref.~\cite{MariavanEgmond2022ATemperature}. We cannot reach too small temperatures numerically, but in the heavy quark limit considered here, given that the transition temperature is way smaller than the quark mass, we find already some precursor of the predicted step function, with plateaux at $0$ or $3$. A more quantitative discussion will be given below, together with what happens above the transition temperature, as well as for ${|\mu|\geq M}$.\footnote{One can already guess from the figure that $\Delta Q_q$ and $\Delta Q_{\bar q}$ approach $0$ in those other situations.} We have checked that the same low-temperature behavior is obtained using the models of Refs.~\cite{Ratti:2005jh, Lo:2013hla, MariavanEgmond2022ATemperature}.\footnote{We mention that some of these models are not based directly on the Polyakov loop potential but, rather, on an effective potential for the gauge-field average in some appropriate gauge. In Appendix~{\ref{app:alternative}}, we argue that similar results as those discussed in the main text apply to this alternative framework.} It is tempting to interpret this universal feature as follows.

At low (zero) temperature, adding a static color charge in the bath is only possible at a huge energy cost, as measured by the smallness of the corresponding Polyakov loop, which causes a significant rearrangement of the medium, screening the unwanted isolated color charge through either a meson-like (${\Delta Q_q+1=0}$ or ${\Delta Q_{\bar q}-1=0}$) or a baryon-like (${\Delta Q_q+1=3}$ or ${\Delta Q_{\bar q}-1=-3}$) configuration, depending on which of these is energetically favored. Actually, the relevant configurations are those that minimize $H-\mu Q$ and not $H$. Then, when bringing a quark probe into a medium with ${\mu<0}$, meson-like states are favored against baryon-like states since they minimize both $H$ and $-\mu Q$. For ${\mu>0}$, in contrast, while meson-like states still minimize $H$, baryon-like states minimize $-\mu Q$. The outcome results then in this case from a compromise between lowering $H$ or lowering $-\mu Q$. For slightly positive $\mu$, meson-like states are favored by continuity with the ${\mu<0}$ case, which we interpret by saying that, as long as the excess of quarks is not too large in the bath, the quark probe can still easily find an antiquark of the medium to form a meson. For too large $\mu$, however, the excess of quarks will be such that, out of number, it will be more likely for the quark probe to find two other quarks than a single antiquark, and baryons will form instead. In equations, forming a meson-like state corresponds to ${H-\mu Q\sim M+\mu}$, while forming a baryon-like state corresponds to ${H-\mu Q\sim 2M-2\mu}$. By equating the two, we find ${\mu=M/3}$ and thus meson-like configurations are favored for ${\mu<M/3}$ while baryon-like configurations are favored for ${\mu>M/3}$. Similar conclusions hold for an antiquark probe upon changing ${\mu\to-\mu}$. In this case, in particular, the meson-like configuration is always favored for ${\mu\ge0}$ (and ${\mu<M}$).

It might seem surprising at first sight that the net quark number gained by the system can be anything other than the quark number of the probe, owing to net quark number conservation. One should keep in mind, however, that the net quark number of the bath is only fixed on average. Upon bringing a probe into the bath, the ensemble distribution of the bath is modified. This implies a change in the average net quark number of the bath, which we have evaluated here. We also note that the value of the Polyakov loop potential $V(\ell,\bar\ell)$ at the relevant extremum gives access to the free-energy density of the bath in the absence of an external probe, but taking into account the feedback from the order parameter. A $\mu$-derivative then gives access to the net quark number density of the bath in the absence of a probe\footnote{The prefactor $2\times 3\times N_f$ accounts for the number of spin states, colors, and flavors.}
\beq
2\times 3\times N_f\times\int \frac{d^3q}{(2\pi)^3}\,\big[n(q)-\bar{n} (q)\big],\nonumber
\eeq
expressed as the difference of momentum-integrated quark and antiquark distribution functions modified by the order parameter:
\beq
n(q)=\frac{1+2\bar\ell e^{\beta(\varepsilon_q-\mu)}+\ell e^{2\beta(\varepsilon_q-\mu)}}{1+3\bar\ell e^{\beta(\varepsilon_q-\mu)}+3\ell e^{2\beta(\varepsilon_q-\mu)}+e^{3\beta(\varepsilon_q-\mu)}},\nonumber
\eeq
and similarly for $\bar n(q)$ upon applying the transformation ${\mu\to-\mu}$ and ${\ell\leftrightarrow\bar\ell}$, see for instance Ref.~\cite{Fu2015RelevanceFluctuations}. The form of these distributions in the limit of vanishing Polyakov loops bears some resemblance to the right-hand side expressions of Eqs.~\eqref{eq:nice}-\eqref{eq:nice2}. Let us stress, however, that this limit is too naive due to the exponentially growing factors that compensate for the smallness of the Polyakov loops in the low temperature range, see in particular Eq.~\eqref{eq:l} and the discussion in Ref.~\cite{Fu2015RelevanceFluctuations}. Moreover, the above quark and antiquark distributions are very different from $\Delta Q_q+1$ and $\Delta Q_{\bar q}-1$. In particular, the former do not discriminate between bringing a quark or an antiquark into the bath. In contrast, $\Delta Q_q+1$ and $\Delta Q_{\bar q}-1$, which are not distributions, measure actual net quark number surpluses as a quark or an antiquark probe is brought into the bath.

Extending the previous analysis to SU($N_c$) gives
\beq
\Delta Q_{q/\bar{q}}\pm 1\simeq\frac{\pm N_c}{1+e^{\mp N_c\beta\mu} f_{\beta M}/f_{(N_c-1)\beta M}}\,.\label{eq:N}
\eeq
Details will be given elsewhere, together with an extension to color probes in other representations. Here, we just stress that, for ${\mu\geq 0}$ (and ${\mu<M}$), bringing an antiquark into the bath always leads the latter to provide a quark, ${\Delta Q_{\bar q}-1=0}$, which we interpret as the system forming a meson-like configuration to screen the color probe. In contrast, bringing a quark into the bath leads the latter either to provide an antiquark ${\Delta Q_q+1=0}$, which we interpret as the system forming a meson-like configuration, or to provide $N_c-1$ quarks, ${\Delta Q_q+1=N_c}$, which we interpret as the system forming a baryon-like configuration. From Eq.~\eqref{eq:N}, the value of $\mu$ above which forming a baryon-like configuration becomes more favorable is found to be ${\mu=(1-2/N_c)M}$. This is easily understood from similar energetic considerations as above and generalizes the formula ${\mu=M/3}$ obtained for ${N_c=3}$.\footnote{For ${N_c=2}$, one finds ${\mu=0}$. This is expected since the representations $2$ and $\bar{2}$ are equivalent, and thus, so are the SU(2) meson- or baryon-like configurations from the color point of view.}

For later use, we finally point out that the estimates (\ref{eq:l}) for $\ell$ and $\bar\ell$ lead to the ratio
\beq
\frac{\bar\ell}{\ell}\simeq\frac{e^{\beta\mu} f_{\beta M}+e^{-2\beta\mu} f_{2\beta M}}{e^{-\beta\mu} f_{\beta M}+e^{2\beta\mu} f_{2\beta M}}\,.\label{eq:ratio}
\eeq
In the limit ${T\to 0}$, this ratio behaves as $e^{-\beta\mu}f_{\beta M}/f_{2\beta M}$ for ${M/3<\mu<M}$, as $e^{2\beta\mu}$ for $|\mu|<M/3$, and as $e^{-\beta\mu}f_{2\beta M}/f_{\beta M}$ for ${-M<\mu<-M/3}$. Thus we find that $\bar\ell$ dominates over $\ell$ for ${0<\mu<M}$ whereas the opposite occurs for ${-M<\mu<0}$. This is a well known result which, in terms of free-energies means for instance that ${\Delta F_{\bar q}<\Delta F_q}$ for ${0<\mu<M}$: in the presence of an excess of quarks in the medium, it is simpler to bring in an antiquark probe. Again, this can be understood in simple terms: the antiquark probe just needs to create a meson-like state with the particles of the medium which is simple since there are more quarks for ${\mu\geq 0}$, on the contrary, the quark probe creates a meson-like state for not too large $\mu$ despite the fact that there are more quarks in the medium, or a baryon-like state if $\mu$ is large enough. Let us stress once more that, just like Eqs.~(\ref{eq:nice}) and (\ref{eq:nice2}), the estimate (\ref{eq:ratio}) is based on a linear approximation which is not fully accurate. A more quantitative description will be given below.

\subsection{${|\mu|>M}$}\label{Xsec4}
The case ${|\mu|>M}$ is not as straightforward as ${|\mu|<M}$. At least in the case of real QCD, we expect di-quarks and color superconductivity to play a role, which is clearly not accounted for by the expression (\ref{eq: quark}) for the matter contribution to the Polyakov loop potential. It would be interesting to see, for instance, using model modifications of (\ref{eq: quark}) that account for color superconductivity, whether $\Delta Q_q$ and $\Delta Q_{\bar q}$ can probe the formation of di-quarks. This is, however, beyond the scope of the present work, and, in what follows, we stick to (\ref{eq: quark}).

This is not a purely academic exercise, though. As we will argue, with a matter contribution given by (\ref{eq: quark}), the region of the phase diagram corresponding to ${T\to 0}$ and ${|\mu|>M}$ can be interpreted as a deconfined phase. Concomitantly, the observables $\Delta Q_q$ and $\Delta Q_{\bar q}$ will be found to vanish, in line with the interpretation that, in the deconfined phase, it should be possible to bring a color probe without significantly affecting the net quark number of the medium.

Although these conclusions will again turn out to be model-independent, their derivation will depend on the model for $V_{\rm glue}$. This is because the integrals in the right-hand side of Eqs.~(\ref{eq:g1}) and (\ref{eq:g2}) are not exponentially suppressed in the case ${|\mu|>M}$, but only polynomially suppressed. The reason is that, even though one can always massage the integrands such that the exponentials have a negative argument, these exponentials cannot be uniformly bounded by a $q$-independent, exponentially vanishing contribution. Indeed, the integration range contains the point ${q=\sqrt{\mu^2-M^2}}$ at which the argument of these exponentials vanishes. Take Eq.~(\ref{eq:g1}) for instance and rewrite it for convenience as
\beq
 \frac{\partial V_{\rm glue}}{\partial\ell} & \!\!\!\!=\!\!\!\! & \frac{3TN_f}{\pi^2}\int_0^\infty dq\,q^2\nonumber\\
&  & \times\,\Bigg\{\frac{e^{-\beta(\varepsilon_q-\mu)}}{1+3\ell e^{-\beta(\varepsilon_q-\mu)}+3\bar\ell e^{-2\beta(\varepsilon_q-\mu)}+e^{-3\beta(\varepsilon_q-\mu)}}\nonumber\\
& & \hspace{0.3cm}+\,\frac{e^{\beta(\varepsilon_q+\mu)}}{1+3\ell e^{\beta(\varepsilon_q+\mu)}+3\bar\ell e^{2\beta(\varepsilon_q+\mu)}+e^{3\beta(\varepsilon_q+\mu)}}\Bigg\},\label{eq:dl}
\eeq
where we have used the identity
\begin{equation}
\frac{e^{-2x}}{1+3\bar\ell e^{-x}+3\ell e^{-2x}+e^{-3x}}=\frac{e^{x}}{1+3\ell e^{x}+3\bar\ell e^{2x}+e^{3x}}\,.\label{eq:id}
\end{equation}
Out of the two types of exponentials appearing in the above expression, only one is uniformly bounded by an exponentially vanishing bound that leads to an exponentially vanishing contribution to the integral as ${T\to 0}$. Neglecting this contribution, we are left with
\beq
0 & \!\!\!\!=\!\!\!\! & \frac{\partial V_{\rm glue}}{\partial\ell}-\frac{T N_f}{\pi^2}\int_0^\infty dq\,q^2\nonumber\\
& & \times\,\frac{3e^{-\sigma\beta(\varepsilon_q-|\mu|)}}{1\!+\!3\ell e^{-\sigma\beta(\varepsilon_q-|\mu|)}\!+\!3\bar\ell e^{-2\sigma\beta(\varepsilon_q-|\mu|)}\!+\!e^{-3\sigma\beta(\varepsilon_q-|\mu|)}}\,,\label{eq:dl2}
\eeq
where $\sigma$ denotes the sign of $\mu$. Consider now the change of variables ${x=\sigma\beta(\varepsilon_q-|\mu|)}$, with ${dx=\beta qdq/(Tx+\mu)}$. The equation becomes
\beq
\frac{\partial V_{\rm glue}}{\partial\ell} & \!\!\!\!\simeq\!\!\!\! & \frac{3T^2N_f}{\pi^2}\int_{\sigma\beta(M-|\mu|)}^{\sigma\infty} \!\!\!\!\!\! dx\,(Tx+\mu)\sqrt{(Tx+\mu)^2-M^2}\nonumber\\
& & \hspace{1.7cm}\times\,\frac{ e^{-x}}{1+3\ell e^{-x}+3\bar\ell e^{-2x}+e^{-3x}}\,.\label{eq:manip}
\eeq
Because ${|\mu|>M}$, the lower boundary goes to $-\sigma\infty$  as ${T\to 0}$, but we note that the integral remains well defined since its integrand is exponentially suppressed on both ends, owing to Eq.~(\ref{eq:id}). This also implies that the prefactor
\beq
(Tx+\mu)\sqrt{(Tx+\mu)^2-M^2}
\eeq
can be safely expanded as ${T\to 0}$ by assuming that $x$ is fixed. Finally, the factor of $\sigma$ appearing in the boundaries of the integral can be pulled in front of the integral. All together, we arrive at
\beq
\frac{\partial V_{\rm glue}}{\partial\ell} & \!\!\!\!\simeq\!\!\!\! & \frac{3N_f}{\pi^2}T^2|\mu|\,(\mu^2-M^2)^{1/2}\,\nonumber\\
& & \times\,\int_{-\infty}^{\infty} \frac{dx\,e^{-x}}{1\!+\!3\ell e^{-x}\!+\!3\bar\ell e^{-2x}\!+\!e^{-3x}}\,.\label{eq:dl3}
\eeq
Using charge conjugation, the equation for $\bar\ell$ can be obtained upon doing ${\ell\leftrightarrow\bar\ell}$ and ${\mu\to-\mu}$:
\beq
\frac{\partial V_{\rm glue}}{\partial\bar\ell} & \!\!\!\!\simeq\!\!\!\! & \frac{3N_f}{\pi^2}T^2|\mu|\,(\mu^2-M^2)^{1/2}\,\nonumber\\
& & \times\,\int_{-\infty}^\infty \frac{dx\,e^{-x}}{1\!+\!3\bar\ell e^{-x}\!+\!3\ell e^{-2x}\!+\!e^{-3x}}\,.\label{eq:dl4}
\eeq
Alternatively, one can follow the same strategy as before.\footnote{Proceeding in a slightly different way, one can obtain an alternative form where $\ell$ and $\bar\ell$ are interchanged and the numerator features $e^{-2x}$ rather than $e^{-x}$. The two forms are equivalent of course and related by Eq.~\eqref{eq:id} in combination with the change of variables ${x\to -x}$.} Note that the $\mu$-dependent factor is the same in both equations. Together with ${V_{\rm glue}(\ell,\bar\ell)=V_{\rm glue}(\bar\ell,\ell)}$, this implies that, if $(\ell,\bar\ell)$ is a solution of the system (\ref{eq:dl3})--(\ref{eq:dl4}), so is $(\bar\ell,\ell)$. Moreover, if the solution is unique, which we have found in all considered cases at low $T$, then we deduce that ${\bar\ell=\ell}$, which corresponds to ${\Delta F_q\simeq \Delta F_{\bar q}}$. This is, in a sense, a necessary condition for ${|\mu|>M}$ to correspond to a deconfined phase because, in such a phase, it should not matter much whether the probe is a quark or an antiquark.

We have checked numerically that the above asymptotic estimates for the quark contribution in the regime ${|\mu|>M}$ and ${T\to 0}$ are accurate.  As announced, the quark contribution is now only polynomially suppressed as ${T\to 0}$. This makes the discussion less universal than before, as it relies on how strong the assumed power-law behavior of the glue contribution is, compared to the $T^2$-suppression of the quark contribution.  The detailed discussion of the different cases can be found in Appendix~{\ref{app:large_mu}}. The final outcome is that, with the approximation (\ref{eq: quark}), the ${|\mu|>M}$ region can be considered as a deconfined phase, and, in line with this interpretation, we find that $\Delta Q_q$ and $\Delta Q_{\bar q}$ vanish at low temperatures in this region in all cases.

\subsection{${|\mu|=M}$}\label{Xsec5}
For completeness, let us consider the case ${|\mu|=M}$, which is similar but not exactly identical to the case ${|\mu|>M}$. In particular, the lower boundary in the integral of Eq.~(\ref{eq:manip}) is strictly equal to $0$. Moreover, the expansion of the prefactor $(Tx+\mu)\sqrt{(Tx+\mu)^2-M^2}$ leads this time to $M^{3/2}T^{1/2}\sqrt{2|x|}$. All together, this gives
\beq
\frac{\partial V_{\rm glue}}{\partial\ell} & \!\!\!\simeq\!\!\! & \sigma\frac{3N_f}{\pi^2}T^{5/2}M^{3/2}\,\nonumber\\
& & \times\,\int_0^{\sigma\infty} \frac{dx\,\sqrt{2|x|}\,e^{-x}}{1\!+\!3\ell e^{-x}\!+\!3\bar\ell e^{-2x}\!+\!e^{-3x}}\,,
\eeq
with $\sigma$ the sign of $\mu$. Similarly,
\beq
\frac{\partial V_{\rm glue}}{\partial\bar\ell} & \!\!\!\simeq\!\!\! & -\sigma\frac{3N_f}{\pi^2}T^{5/2}M^{3/2}\,\nonumber\\
& & \times\,\int_0^{-\sigma\infty} \frac{dx\,\sqrt{2|x|}\,e^{-x}}{1\!+\!3\bar\ell e^{-x}\!+\!3\ell e^{-2x}\!+\!e^{-3x}}\,.
\eeq
Thus, once more, the discussion of what happens as ${T\to 0}$ depends on the strength of the quark contribution vs the glue contribution.

\section{Non-zero temperatures}\label{Xsec6}

Fig.~\ref{fig: Qqbar of mu} also shows results at non-zero temperatures, obtained using the model of Ref.~\cite{MariavanEgmond2022ATemperature}. Similar results can be obtained with the other glue potentials. We observe in particular that ${\Delta Q_{q/\bar q}\to 0}$ above the deconfinement transition. This is again a universal feature, at least at asymptotically large temperatures, which remains true until near the transition in the heavy-quark regime considered here. In this case also, the plateaux observed as ${T\to 0}$ extend throughout most of the confined phase.

\subsection{Asymptotically large temperatures}\label{Xsec7}
In this limit, we can simply invoke asymptotic freedom. Anticipating some difficulties similar to those encountered in the analysis of the ${|\mu|>M}$ region, see Appendix~{\ref{app:large_mu}}, we work with the related function \cite{Dumitru:2013xna}
\beq
W(\bar r_3,\bar r_8)=V(\ell(\bar r_3,\bar r_8),\bar\ell(\bar r_3,\bar r_8)),
\eeq
with\footnote{Our notation differs slightly from the one in Ref.~\cite{Dumitru:2013xna}.}
\beq
\ell(\bar r_3,\bar r_8)=\frac{1}{3}\sum_\rho e^{i\rho\cdot \bar r}, \quad \bar\ell(\bar r_3,\bar r_8)=\frac{1}{3}\sum_\rho e^{-i\rho\cdot \bar r},
\eeq
where ${\bar r=(\bar r_3,\bar r_8)}$\footnote{For a real chemical potential, $\bar{r}_8$ should be taken purely imaginary, see for instance Ref.~\cite{Reinosa:2019xqq}.} is a two-component vector and $\rho$  runs over the defining weights of SU(3): $(1/2,1/(2\sqrt{3}))$, $(-1/2,1/(2\sqrt{3}))$ and $(0,-1/\sqrt{3})$.

The renormalization group, together with dimensional analysis, allows one to write
\beq
& & W(\bar r;T,\mu;s_0,g(s_0),M(s_0))\nonumber\\
& & \hspace{1.0cm}=\,W(\bar r_3,\bar r_8;T,\mu;T,g(T),M(T))\nonumber\\
& & \hspace{1.0cm}=\,T^4W(\bar r_3,\bar r_8;1,\mu/T;1,g(T),M(T)/T)\,.\ 
\eeq
Thus, for large temperatures, one can use perturbation theory, and, at leading order, one is led to consider the Weiss potential
\beq
& & W(\bar r_3,\bar r_8;T,\mu;s_0,g(s_0),M(s_0))\nonumber\\
& & \hspace{0.7cm}\simeq\,T^4w_{\rm Weiss}(\bar r_3,\bar r_8;\mu/T)\nonumber\\
& & \hspace{0.7cm}\simeq\,T^4\left[w_{\rm Weiss}(\bar r_3,\bar r_8;0)+\frac{\mu}{T}w^{(0,0,1)}_{\rm Weiss}(\bar r_3,\bar r_8;0)\right].
\eeq
Since the extremum of $w_{\rm Weiss}(\bar r_3,\bar r_8;0)$ is located at ${\bar r=(0,0)}$, we deduce that the extremum $\bar r(T)$ of the complete potential obeys
\beq
\bar r(T)=a(T)+\frac{\mu}{T}b(T)+{\cal O}\left(\frac{\mu^2}{T^2}\right),
\eeq
with ${a(T)\to 0}$. The Polyakov loop is then
\beq
\ell= \ell(a)+ \frac{\mu}{T}\frac{b}{3}\cdot\sum_\rho \rho\,e^{i\rho\cdot a}+{\cal O}\left(\frac{\mu^2}{T^2}\right),
\eeq
and similarly for $\bar\ell$ with ${\mu\to -\mu}$. It follows that
\beq
\Delta Q_{q/\bar q}\simeq\frac{\pm\frac{b}{3}\cdot\sum_\rho \rho\,e^{\pm i\rho\cdot a}}{1\pm\frac{\mu}{T}\frac{b}{3}\cdot\sum_\rho \rho\,e^{\pm i\rho\cdot a}},
\eeq
which approach $0$ as ${T\to 0}$ since ${a\to 0}$ and ${\sum_\rho \rho=0}$.

\subsection{Nearing the transition from above}\label{Xsec8}
In the heavy-quark regime, the previous result extends down to near the deconfinement transition. The reason is that $T_c(\mu)\leq T_c(\mu=0)\lesssim T_c^{\rm YM}\ll M$ and so one can employ similar approximations as before. The only difference is that the linearization of the equations needs to be done around $\ell_{\rm YM}(T)$ and $\bar\ell_{\rm YM}(T)$ at that temperature. Suppose first that $T$ is slightly above $T_c^{\rm YM}$, and then close to $T_c(\mu=0)$. One then finds that ${\ell=\ell_{\rm YM}+\delta\ell}$ and ${\bar\ell=\bar\ell_{\rm YM}+\delta\bar\ell}$, with $\delta\ell$ and $\delta\bar\ell$ given by a linear combination of the four exponentials appearing in Eqs.~(\ref{eq:gap1}) and (\ref{eq:gap2}) which are all exponentially suppressed, while $\ell_{\rm YM}$ and $\bar\ell_{\rm YM}$ are not. Moreover, since the latter do not depend on $\mu$, one finds
\beq
\Delta Q_q\simeq\frac{T}{\ell_{\rm YM}}\partial_\mu\delta\ell\simeq 0, \quad \Delta Q_{\bar q}\simeq\frac{T}{\bar\ell_{\rm YM}}\partial_\mu\delta\bar\ell\simeq 0\,.
\eeq
These conclusions extend even below the line ${T=T_c^{\rm YM}}$ and down to close to the curve $T_c(\mu)$. The point is that, above this curve, while $\ell$ and $\bar\ell$ are not exponentially suppressed, $\partial_\mu\ell$ and $\partial_\mu\bar\ell$ are.

We mention that this discussion, together with the one for ${|\mu|>M}$, shows that $\Delta Q_{q/\bar q}$ behave as order parameters for the deconfinement transition since they are essentially $0$ in the deconfined phase, while they take non-zero values in the confined phase.

\subsection{Confined phase}\label{Xsec9}
We also find that the plateaux for ${\Delta Q_q+1}$ that we found in the limit ${T\to 0}$ in the region ${0<\mu<M}$ extend throughout most of the confined phase in the heavy-quark case. There, they are connected by a smooth transition with ${0<\Delta Q_q+1<3}$, which we interpret as a thermal competition between the formation of a meson-like or a baryon-like configuration. The way the transition occurs from $0$ to $3$ depends on the model for $V_{\rm glue}$, but the presence of plateaux at finite $T$ is again a model-independent feature related to the fact that the transition temperature lies far below the considered heavy-quark masses. To make our argument rigorous, though, we need to go beyond the linear approximation used above.

In fact, because ${\partial^2_\ell V_{\rm glue}=\partial^2_{\bar \ell} V_{\rm glue}=0}$ at ${(\ell,\bar\ell)=(0,0)}$, each linearized equation around ${(\ell,\bar\ell)=(0,0)}$ involves only $\ell$ or $\bar\ell$, and one cannot exclude that these terms are respectively of the same order as $\bar\ell^2$ and $\ell^2$. Actually, the linearized equations lead to the estimate (\ref{eq:l}) which implies that, in the limit ${T\to 0}$, the ratio $\bar\ell^2/\ell$ behaves as $f^2_{\beta M}/f_{2\beta M}$ for ${M/3<\mu<M}$, as $e^{3\beta\mu}f_{\beta M}$ for $|\mu|<M/3$, and as $e^{-3\beta\mu}f^2_{2\beta M}/f_{\beta M}$ for ${-M<\mu<-M/3}$. In particular, it is not exponentially suppressed in the region $M/3<\mu<M$. Thus, in the region ${0<\mu<M}$, a more consistent analysis should include the term $\frac{1}{2}\partial_{\bar\ell}^3V_{\rm glue} \times(0,\bar\ell^2)^{\rm t}$ in the left-hand side of Eq.~\eqref{eq:gap3}.

While the equation fixing $\bar\ell$ is not changed, leading to the same result as before:
\beq
\bar\ell\simeq\frac{C}{\partial_\ell\partial_{\bar\ell} V_{\rm glue}}(
e^{\beta\mu} f_{\beta M}+e^{-2\beta\mu} f_{2\beta M}),
\eeq
and similarly for $\Delta Q_{\bar q}-1$, the equation for $\ell$ becomes
\beq
\partial_\ell\partial_{\bar\ell}V_{\rm glue}\ell+\frac{1}{2}\partial_{\bar\ell}^3V_{\rm glue} \bar\ell^2=C(e^{-\beta\mu} f_{\beta M}+e^{2\beta\mu} f_{2\beta M}),
\eeq
and thus
\beq
\ell & \!\!\!\!\simeq\!\!\!\! & \frac{C}{\partial_\ell\partial_{\bar\ell}V_{\rm glue}}\Bigg[e^{-\beta\mu} f_{\beta M}+e^{2\beta\mu} f_{2\beta M}-\frac{C}{2}\frac{\partial_{\bar\ell}^3V_{\rm glue}}{(\partial_\ell\partial_{\bar\ell}V_{\rm glue})^2}\nonumber\\
&  & \times\,(e^{2\beta\mu} f^2_{\beta M}+2e^{-\beta\mu}f_{\beta M}f_{2\beta M}+e^{-4\beta\mu} f^2_{2\beta M})\Bigg]\,.
\eeq
The last exponential in the second line can be clearly neglected, and for later convenience, we rewrite the resulting expression as
\beq
\ell & \!\!\!\!\simeq\!\!\!\! & \frac{C}{\partial_\ell\partial_{\bar\ell}V_{\rm glue}}\Big[D_1e^{-\beta\mu} f_{\beta M}+D_2e^{2\beta\mu} f_{2\beta M}\Big],
\eeq
with\footnote{Note that the second term in $D_1$ is exponentially suppressed so it could be safely neglected. On the contrary, the second term in $D_2$ is only power-law suppressed and plays a more important role in the region where $\beta M$ is large but not arbitrarily large. The reason why we have kept both corrections is that they are the only ones that modify the prefactors in front of $e^{-\beta\mu}$ and $e^{2\beta\mu}$ in the expression for $\ell$. Higher corrections involve other powers of $e^{\beta\mu}$.}
\beq
D_1 & \!\!\!\!=\!\!\!\! & 1-C\frac{\partial^3_{\bar\ell}V_{\rm glue}}{(\partial_\ell\partial_{\bar\ell}V_{\rm glue})^2}f_{2\beta M},\\
D_2 & \!\!\!\!=\!\!\!\! & 1-\frac{C}{2}\frac{\partial^3_{\bar\ell}V_{\rm glue}}{(\partial_\ell\partial_{\bar\ell}V_{\rm glue})^2}\frac{f^2_{\beta M}}{f_{2\beta M}},
\eeq
and where it is again implicitly understood that the derivatives of $V_{\rm glue}$ are evaluated at ${\ell=\bar\ell=0}$.

\begin{figure}
    \centering
    \includegraphics[width=.9\linewidth]{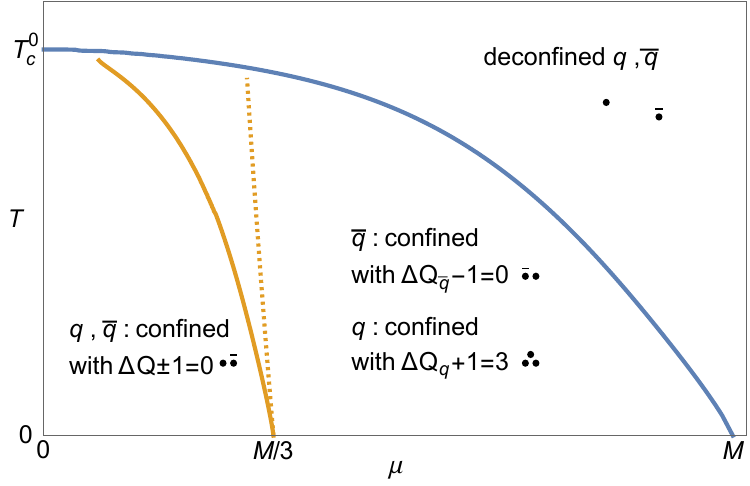}
    \caption{Phase diagram of heavy-quark QCD, as resulting from Eqs.~(\ref{eq:V}) and (\ref{eq: quark}), with $V_{\rm glue}(\ell,\bar\ell)$ modeled as in Ref.~\cite{MariavanEgmond2022ATemperature}. The outer line shows the confinement-deconfinement transition.
    The line within the confined phase separates the regions where the medium screens the quark probe $q$ via a meson-like ($\to0$) or a baryon-like ($\to3$) configuration. It was cut around where the plateaux disappear. The dashed line is the qualitative estimate derived from Eq.~\eqref{eq:nice}, the quantitative estimate from Eq.~\eqref{eq:corr} is indistinguishable from the full result.}
    \label{fig: phase-dia}
\end{figure}

Then
\beq
\Delta Q_q=\frac{-D_1e^{-\beta\mu} f_{\beta M}+2D_2e^{2\beta\mu} f_{2\beta M}}{D_1e^{-\beta\mu} f_{\beta M}+D_2e^{2\beta\mu} f_{2\beta M}}
\eeq
and thus
\beq\label{eq:corr}
    \Delta Q_q+1\simeq\frac{3}{1+D\, e^{-3\beta\mu}f_{\beta M}/f_{2\beta M}},
\eeq
which is similar to Eq.~\eqref{eq:nice} but with a temperature- and model-dependent correction factor ${D=D_1/D_2}$. Owing to the definition of $C$, see Eqs.~\eqref{eq:gap1}-\eqref{eq:gap2}, and the assumption that $V_{\rm glue}$ behaves as a power law as ${T\to 0}$, one finds that $D$ goes to a constant or vanishes as a power law. This implies that in the limit ${T\to 0}$, one retrieves a step function separating $0$ and $3$ at ${\mu=M/3}$, in line with our findings obtained using the linear approximation.\footnote{The convergence towards the step function is not uniform around ${\mu=M/3}$ if ${D<0}$. We have checked that ${D>0}$ in all the considered models, ensuring a uniform limit. This sign might be rooted in the first-order nature of the SU(3) pure-gauge transition, which favors ${\partial^3_{\bar\ell}V_{\rm glue}<0}$.}

 But Eq.~\eqref{eq:corr} also implies that the two plateaux at $0$ and $3$ extend to most of the confined phase, as anticipated. Defining the location of the change from $0$ to $3$ from the inflection of ${\Delta Q_q+1}$, we find the value
 \beq
 \mu=(T/3)\ln\,(D\,f_{\beta M}/f_{2\beta M}),\label{eq:estimate}
 \eeq
which depends on $V_{\rm glue}$ through $D$. This defines a curve that can be represented in the phase diagram of heavy-quark QCD, see Fig.~\ref{fig: phase-dia}, together with the confinement-deconfinement transition curve. Using the model of Ref.~\cite{MariavanEgmond2022ATemperature} as an example, we have checked that the estimate (\ref{eq:estimate}) agrees pretty well with the full numerical determination.

We stress that the above description is valid as long as we do not get too close to the transition so that the Polyakov loops remain small enough. Close to the transition, particularly near a critical point, other interesting effects occur, such as $\Delta Q_q+1$ exceeding $3$, whose discussion we leave for a future investigation.

\section{Conclusions and outlook}\label{Xsec10}
In this Letter, we have proposed new theoretical observables as probes of the quark content of the relevant degrees of freedom in the confined and deconfined phases of QCD. We have tested these observables in the heavy-quark regime, where there is a clear distinction between such phases. We have found that these observables, which count the net quark number gained by the system upon bringing a probe, while they coincide with the net quark number of the probe in the deconfined phase, acquire rather different values in the confined, low temperature phase, which turn out to be multiples of $3$, in line with the picture that the probe is absorbed into mesonic or baryonic states.

The present analysis does not tell us about the color representations formed to screen the color probes, so we cannot yet test the standard expectation that they correspond to color-neutral states. This is why we have referred to meson-like or baryon-like configurations rather than actual mesons or baryons. One way to refine the analysis would be to evaluate the expectation value for the Casimir operator associated with the color charge. More generally, evaluating thermal averages of conserved charges in the presence of the Polyakov loop could provide fruitful ways to scan the QCD phase diagram. It would also be interesting to confront the present results with simulations in the heavy-quark regime \cite{Fromm2012ThePotentials}.

Our results were derived in the heavy-quark regime under the assumption that the one-loop quark contribution provides a good approximation of the matter sector. We should, of course, test our results against the inclusion of higher-order corrections. We have already started exploring the impact of the two-loop corrections. As already indicated in the main text, an even more stringent test would be to use the results of numerical simulations in heavy-quark QCD \cite{deForcrand:2010he}. We believe that our results could also partly extend to the physical QCD case. In particular, working in the Landau gauge and exploiting the well-tested expansion in the inverse number of colors and the fact that the pure glue coupling is not that large \cite{Duarte:2016iko, Reinosa:2017qtf}, the quark contribution to the Polyakov loop potential is given by an effective one-loop contribution involving the rainbow-resummed quark propagator. At low temperatures, we expect this loop to be dominated by low momenta and thus by the constituent quark mass, as fixed by chiral symmetry breaking. Then, the $\mu$-profile of the net quark number gained by the system at low enough temperatures should not be altered. We have confirmed this expectation in a model calculation coupling the glue potential of Ref.~\cite{MariavanEgmond2022ATemperature} to a Nambu-Jona-Lasinio model for the quark sector. Results will be reported elsewhere. A more ambitious calculation would involve coupling the glue potential to the rainbow-resummed quark contribution.

In summary, we have identified an observable sensitive to the quark versus hadron content of a bath of quarks and gluons. This opens interesting perspectives as it could be used, for instance, to investigate the persistence of chirally symmetric, hadron-like states above the chiral phase transition, see e.g.~\cite{Glozman2022ChiralDiagram}, the location of critical end-points, Cooper pairing at large $\mu$, or the formation of exotic hadron states beyond the usual mesons or baryons.

\section*{Acknowledgements}

We would like to thank {J.~P. Blaizot,} {C.~Lorc\'e,} {J.~M.~Pawlowski,} {J.~Serreau,} {M.~Tissier,} and {N.~Wschebor} for useful suggestions on the manuscript.

\section*{Data availability}
No data was used for the research described in the article.

\section*{Declaration of interests}
The authors declare that they have no known competing financial
interests or personal relationships that could have appeared to
influence the work reported in this paper.

\appendix

\def\thefigure{A.\arabic{figure}}
\setcounter{figure}{0}
\section{Polyakov loop potential and alternatives}\label{app:alternative}
The analysis in the main text relies on a few, general and well established properties of the Polyakov loop potential $V(\ell,\bar\ell)$ (specific form of its matter contribution for heavy quark masses and confining glue contribution at low temperatures) plus one more specific feature which seems to be shared by most successful descriptions of confinement (the glue contribution dominates over the matter one for ${T\to 0}$ and ${|\mu|<M}$). This implies, in particular, that, in any approach that models the Polyakov loop potential while respecting these few properties, the behaviors of the Polyakov loops (as derived from the extremization of the Polyakov loop potential), and, thus, of the net quark number gains, should comply with those described in this work. References \cite{Ratti:2005jh} and \cite{Lo:2013hla} provide examples of such models, and we have checked that they indeed lead to the expected behaviors.

There exist other ways to access the Polyakov loop, however, that do not rely directly on the Polyakov loop potential, but, rather, on the evaluation of a gauge-field effective potential in some chosen gauge. The gauge-field expectation value obtained by extremizing this potential is then used to evaluate the Polyakov loop. Let us now argue that the results of the present work extend to this framework as well and are equally robust to the modeling of the gauge-field effective potential, provided certain features are preserved by the modeling.

\begin{figure}
    \centering
    \includegraphics[width=.8\linewidth]{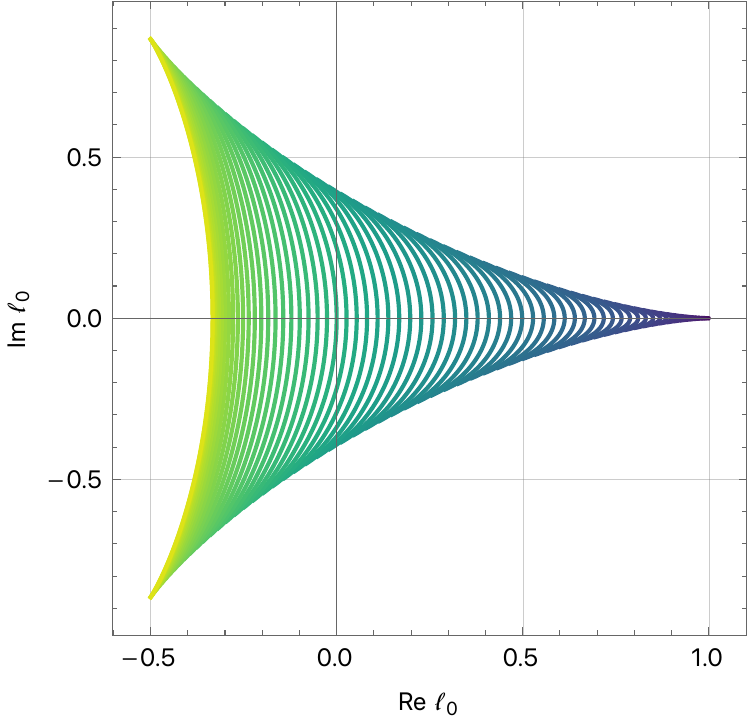}
    \caption{For a grid of values of $r_3$ between $0$ (dark) and $2\pi$ (light), we show $({\rm Re}\,\ell_0,{\rm Im}\,\ell_0)$ as $r_8$ is varied such that $(r_3,r_8)$ remains in the same Weyl chamber. The corresponding curves do not cross each other and cover a certain region in the plane $({\rm Re}\,\ell_0,{\rm Im}\,\ell_0)$.}
    \label{fig:lrli}
\end{figure}

When it comes to studying confinement vs deconfinement, a rather popular gauge is the background Landau gauge or Landau-deWitt gauge \cite{Braun:2007bx}. Rather than just one gauge, this is actually a class of gauges parametrized by a gluonic background. In the SU(3) case, where the only fundamental Polyakov loops are $\ell$ and $\bar\ell$, it is pretty convenient to choose constant and temporal backgrounds along the Abelian directions of the algebra. Such backgrounds are parametrized by two numbers $\bar r_3$ and $\bar r_8$, and, for any possible choice, one has then access to the effective potential $V_{\bar r_3,\bar r_8}(r_3,r_8)$ for the gauge-field averages $r_3$ and $r_8$ of the temporal and Abelian components of the gauge-field.

Now, it is easily seen that, if we assume once again a one-loop form of the matter contribution to $V_{\bar r_3,\bar r_8}(r_3,r_8)$, the latter does not depend on $\bar r_3$ or $\bar r_8$, and writes identically as
\beq
V^{\rm quark}_{\bar r_3,\bar r_8}(r_3,r_8)=V_{\rm quark}(\ell_0(r_3,r_8),\bar\ell_0(r_3,r_8)),
\eeq
see for instance Ref.~\cite{Reinosa:2019xqq}, where $V_{\rm quark}(\ell,\bar\ell)$ is given in Eq.~(\ref{eq: quark}) and
\beq
\ell_0(r_3,r_8) & \!\!\!\!=\!\!\!\! & \frac{e^{-i\frac{r_8}{\sqrt{3}}}+2e^{i\frac{r_8}{2\sqrt{3}}}\cos(r_3/2)}{3},\label{eq:l0}\\
\bar\ell_0(r_3,r_8) & \!\!\!=\!\!\! & \frac{e^{i\frac{r_8}{\sqrt{3}}}+2e^{-i\frac{r_8}{2\sqrt{3}}}\cos(r_3/2)}{3}\,.\label{eq:lb0}
\eeq
By inverting the relation between $(r_3,r_8)$ and $(\ell_0,\bar\ell_0)$, one can enforce the same type of identity for the glue contribution\footnote{This is more subtle than it first meets the eye, see the discussion below. Note also that we use the same notation $V^{\rm glue}_{\bar r_3,\bar r_8}$ to denote two distinct functions, which are identified by their arguments, either $r_3$ and $r_8$, or $\ell_0$ and $\bar\ell_0$.}
\beq
V^{\rm glue}_{\bar r_3,\bar r_8}(r_3,r_8)=V^{\rm glue}_{\bar r_3,\bar r_8}(\ell_0(r_3,r_8),\bar\ell_0(r_3,r_8)),\label{eq:A4}
\eeq
so that the extremization of $V_{\bar r_3,\bar r_8}(r_3,r_8)$ is tantamount to the extremization of
\beq
V_{\bar r_3,\bar r_8}(\ell_0,\bar\ell_0)=V^{\rm glue}_{\bar r_3,\bar r_8}(\ell_0,\bar\ell_0)+V_{\rm quark}(\ell_0,\bar\ell_0)\,.\label{eq:V1}
\eeq
Provided the modeling of $V^{\rm glue}_{\bar r_3,\bar r_8}(r_3,r_8)$ implies that $V^{\rm glue}_{\bar r_3,\bar r_8}(\ell_0,\bar\ell_0)$ dominates over the quark contribution as ${T\to 0}$ and ${|\mu|<M}$ and is confining, we can apply the same reasoning as above to deduce the behavior of the ``Polyakov loops'' $\ell_0$ and $\bar\ell_0$ at the relevant extremum of $V_{\bar r_3,\bar r_8}(\ell_0,\bar\ell_0)$, and, thus, of the associated ``net quark number responses''
\beq
\Delta Q_0=T\frac{\partial\ln\ell_0}{\partial\mu} \quad {\rm and} \quad \Delta \bar Q_0=T\frac{\partial\ln\bar\ell_0}{\partial\mu}\,.
\eeq
The dominance property is usually obeyed by known modelings of $V^{\rm glue}_{\bar r_3,\bar r_8}(r_3,r_8)$, see for instance Ref.~\cite{MariSurkau2024DeconfinementDependences}. The confining property, on the other hand, can be ensured by choosing specific values of $\bar r_3$ and $\bar r_8$, referred to as confining and denoted $\bar r_{3c}$ and $\bar r_{8c}$, see Ref.~\cite{MariavanEgmond2022ATemperature}.

\begin{figure}
    \centering
    \includegraphics[width=.8\linewidth]{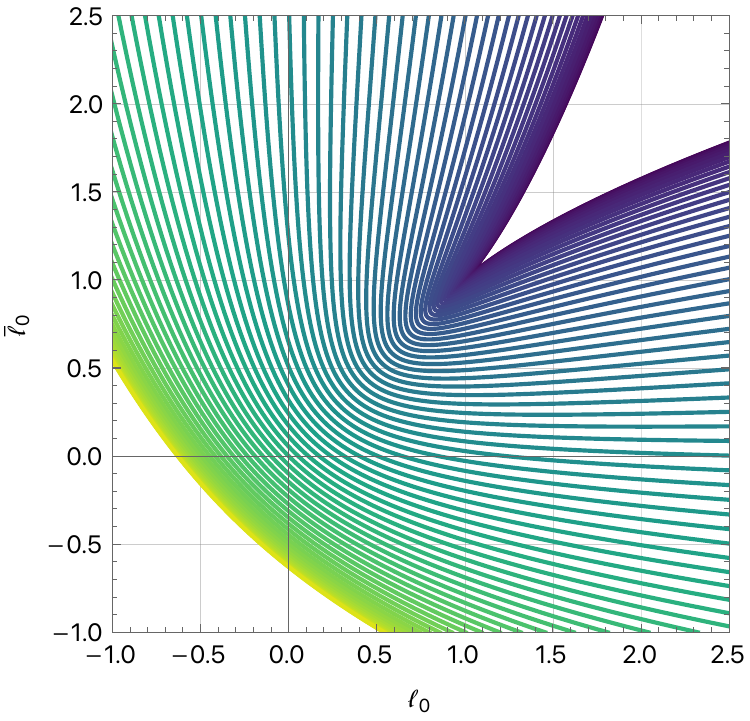}
    \caption{For a grid of values of $\hat r_3$ between $0$ and $2\pi$, we show $(\ell_0,\bar\ell_0)$ as $\hat r_8$ is varied. The corresponding curves do not cross each other and cover a certain region in the plane $(\ell_0,\bar\ell_0)$.}
    \label{fig:llb}
\end{figure}

It is to be stressed, however, that $V_{\bar r_3,\bar r_8}(\ell_0,\bar\ell_0)$ does not correspond to the Polyakov loop potential discussed in the main text. Otherwise stated, $\ell_0(r_3,r_8)$ and $\bar\ell_0(r_3,r_8)$ evaluated at the relevant extremum of $V_{\bar r_3,\bar r_8}(r_3,r_8)$ do not correspond to the physical Polyakov loops. The latter are obtained from some other functions $\ell_{\bar r_3,\bar r_8}(r_3,r_8)$ and $\bar\ell_{\bar r_3,\bar r_8}(r_3,r_8)$ that can in principle be computed within a systematic loop expansion in the considered gauge and of which $\ell_0(r_3,r_8)$ and $\bar\ell_0(r_3,r_8)$ represent the tree-level expressions, see for instance Ref.~\cite{Reinosa:2014zta} for an evaluation of the NLO corrections within the Curci-Ferrari model with ${(\bar r_3,\bar r_8)=(r_3,r_8)}$. However, upon inversion of the relations (\ref{eq:l0}) and (\ref{eq:lb0}), one can deduce the physical Polyakov loops $(\ell,\bar\ell)$ from the knowledge of $(\ell_0,\bar\ell_0)$. This is what we would like to exploit now.

Consider the case ${|\mu|<M}$ and ${T\to 0}$ (the case ${|\mu|\geq M}$ will be treated in the next section). The matter integrals that contribute to the relation between $(\ell_0,\bar\ell_0)$ and $(\ell,\bar\ell)$  can be neglected in this limit. Moreover, for any choice of confining backgrounds, one can show that this ``pure-gauge'' relation obeys the center symmetry property
\beq
\ell(e^{i2\pi/3}\ell_0,e^{-i2\pi/3}\bar\ell_0) & \!\!\!\!=\!\!\!\! & e^{i2\pi/3}\ell(\ell_0,\bar\ell_0),\label{eq:l0center}\\
\bar\ell(e^{i2\pi/3}\ell_0,e^{-i2\pi/3}\bar\ell_0) & \!\!\!=\!\!\! & e^{-i2\pi/3}\bar\ell(\ell_0,\bar\ell_0)\,.\label{eq:lb0center}
\eeq
Now, since $\ell_0$ and $\bar\ell_0$ get exponentially small as ${T\to 0}$, so do $\ell$ and $\bar\ell$, and we have
\beq
\left(
\begin{array}{@{}l@{}}
\ell\\
\bar\ell
\end{array}
\right)\simeq\left(
\begin{array}{@{}ll@{}}
\partial_{\ell_0}\ell & \partial_{\bar\ell_0}\ell\\
\partial_{\ell_0}\bar\ell & \partial_{\bar\ell_0}\bar\ell
\end{array}
\right)\left(
\begin{array}{@{}l@{}}
\ell_0\\
\bar\ell_0
\end{array}
\right),
\eeq
where the components of the square matrix are to be taken at ${\ell_0=\bar\ell_0=0}$. The center symmetry (\ref{eq:l0center}) and (\ref{eq:lb0center}) implies that $\partial_{\ell_0}\bar\ell$ and $\partial_{\bar\ell_0}\ell$ vanish for ${\ell_0=\bar\ell_0=0}$, and so we are left with
\beq
\ell\simeq \partial_{\ell_0}\ell\times\ell_0 \quad {\rm and} \quad \bar\ell\simeq \partial_{\bar\ell_0}\bar\ell\times\bar\ell_0\,.
\eeq
From this, we deduce that
\beq
\Delta Q\simeq \Delta Q_0+T\frac{\partial\ln \partial_{\ell_0}\ell}{\partial\mu},
\eeq
but, since the matter contributions have been neglected, the second term vanishes, and we deduce that ${\Delta Q\simeq\Delta Q_0}$ as ${T\to 0}$, meaning that the behaviors for the net quark number gains as obtained from the gauge-field potential comply with those described in the present work.

Before concluding this section, some additional remarks are in order. First, the inversion of the relations (\ref{eq:l0}) and (\ref{eq:lb0}) is not completely trivial. A simple way to see this is to notice that
\beq
\ell_0(r_3,r_8) & \!\!\!\!=\!\!\!\! & \frac{1}{3}\sum_\rho e^{i\rho\cdot r},\\
\bar\ell_0(r_3,r_8) & \!\!\!=\!\!\! & \frac{1}{3}\sum_\rho e^{-i\rho\cdot r},
\eeq
where ${r\equiv (r_3,r_8)}$ and the label $\rho$ runs over the defining weights of SU(3): $(1/2,1/(2\sqrt{3}))$, $(-1/2,1/(2\sqrt{3}))$ and $(0,-1/\sqrt{3})$. In this form, it is easily checked that $\ell_{0}(r_3,r_8)$ and $\bar\ell_0(r_3,r_8)$ are invariant under the following transformations:
\beq
r\to r-2(r\cdot\alpha)\,\alpha \quad {\rm and} \quad r\to r+4\pi\alpha,\label{eq:transfos}
\eeq
where $\alpha$ denotes any of the SU(3) roots, obtained as non-vanishing differences of the weights listed above: $\pm (1,0)$, $\pm(1/2,\sqrt{3}/2)$ and $\pm(1/2,-\sqrt{3}/2)$.

To exploit the above transformations, it is common to distinguish two situations. In the presence of an imaginary chemical potential, it can be argued that one should consider $r_3$ and $r_8$ real, in which case $\ell_0$ and $\bar\ell_0$ are complex, and actually complex conjugate of each other. It is then more convenient to think of the mapping (\ref{eq:l0}) and (\ref{eq:lb0}) as a mapping from $(r_3,r_8)$ to $({\rm Re}\,\ell_0,{\rm Im}\,\ell_0)$:
\beq
{\rm Re}\,\ell_0
& \!\!\!\!=\!\!\!\! & \frac{\cos\,(r_8/\sqrt{3})+2\cos\,(r_8/(2\sqrt{3}))\cos\,(r_3/2)}{3},\nonumber\\
{\rm Im}\,\ell_0
& \!\!\!\!=\!\!\!\! & \frac{-\sin\,(r_8/\sqrt{3})+2\sin\,(r_8/(2\sqrt{3}))\cos\,(r_3/2)}{3}\,.\nonumber
\eeq
The transformations (\ref{eq:transfos}) subdivide the plane of $r$ into triangular regions, known as Weyl chambers, connected to each other via reflections with respect to their edges and such that two points connected by such a reflection have the same value of ${\rm Re}\,\ell_0$ and ${\rm Im}\,\ell_0$. On the other hand, it can be checked that, when restricted to a given Weyl chamber, the mapping from $(r_3,r_8)$ to $({\rm Re}\,\ell_0,{\rm Im}\,\ell_0)$ is invertible, see Fig.~\ref{fig:lrli} for an illustration. Note that not all values of $({\rm Re}\,\ell_0,{\rm Im}\,\ell_0)$ are reached in $\mathds{R}^2$. Moreover, due to the symmetry identities, the Jacobian of the mapping between $(r_3,r_8)$ to $({\rm Re}\,\ell_0,{\rm Im}\,\ell_0)$ vanishes on the edges of the Weyl chambers.

In the presence of a real chemical potential, it can be argued that one should still take $r_3=\hat r_3$ real while promoting $r_8=i\hat r_8$ to a purely imaginary number. In this case, the mapping (\ref{eq:l0}) and (\ref{eq:lb0}) maps two real and independent numbers $\hat r_3$ and $\hat r_8$ onto two real and independent numbers $\ell$ and $\bar\ell$:
\beq
\ell_0 & \!\!\!\!=\!\!\!\! & \frac{e^{\frac{\hat r_8}{\sqrt{3}}}+2e^{-\frac{\hat r_8}{2\sqrt{3}}}\cos(\hat r_3/2)}{3},\\
\bar\ell_0 & \!\!\!=\!\!\! & \frac{e^{-\frac{\hat r_8}{\sqrt{3}}}+2e^{\frac{\hat r_8}{2\sqrt{3}}}\cos(\hat r_3/2)}{3}\,.
\eeq
The only transformations (\ref{eq:transfos}) that are relevant in this case correspond to ${\alpha=\pm(1,0)}$. They subdivide the plane of $\hat r\equiv (\hat r_3,\hat r_8)$ into vertical bands with ${\hat r_3\in\,[2\pi n,2\pi (n+1)]}$. The transformation is invertible when restricted to any of these bands, see Fig.~\ref{fig:llb} for an illustration, but the Jacobian of the mapping between $(\hat r_3,\hat r_8)$ and $(\ell_0,\bar\ell_0)$ vanishes on the edges. Once again, not all possible values of $(\ell_0,\bar\ell_0)$ are reached in $\mathds{R}^2$.

The previous considerations show that there is actually one possible inversion of Eqs.~(\ref{eq:l0}) and (\ref{eq:lb0}) per Weyl chamber or per band, depending on the considered case. It is then not completely clear which inversion should be chosen to define $V^{\rm glue}_{\bar r_3,\bar r_8}(\ell_0,\bar\ell_0)$ in Eq.~(\ref{eq:A4}). However, in the case where $\bar r_3$ and $\bar r_8$ are chosen to be confining, the relevant extremum of $V^{\rm glue}_{\bar r_3,\bar r_8}(r_3,r_8)$ equals the confining background in the confined phase. In the deconfined phase, the relevant extremum deviates from this value; however, in all models we have considered, we find that it remains within the same Weyl chamber as the chosen confining background. It is then over this Weyl chamber that the inversion should be considered for the definition of  $V^{\rm glue}_{\bar r_3,\bar r_8}(\ell_0,\bar\ell_0)$.

The second remark is that an alternative to $V_{\bar r_3,\bar r_8}(r_3,r_8)$ is to consider ${\tilde V(\bar r_3,\bar r_8)\equiv V_{\bar r_3,\bar r_8}(r_3=\bar r_3,r_8=\bar r_8)}$. It can be argued that the extremization of this functional with respect to $\bar r_3$ and $\bar r_8$ is equivalent to the extremization of $V_{\bar r_3,\bar r_8}(r_3,r_8)$ with respect to $r_3$ and $r_8$. As before, the matter contribution in the heavy-quark regime writes
\beq
\tilde V_{\rm quark}(\bar r_3,\bar r_8)=V_{\rm quark}(\ell_0(\bar r_3,\bar r_8),\bar\ell_0(\bar r_3,\bar r_8))\,.
\eeq
Similarly, upon inversion, one can enforce the same type of identity for the glue contribution,
\beq
\tilde V^{\rm glue}(\bar r_3,\bar r_8)=\tilde V_{\rm glue}(\ell_0(\bar r_3,\bar r_8),\bar\ell_0(\bar r_3,\bar r_8)),
\eeq
so that the extremization of $W_{\bar r_3,\bar r_8}(\bar r_3,\bar r_8)$ is tantamount to the extremization of
\beq
\tilde V(\ell_0,\bar\ell_0)\equiv \tilde V^{\rm glue}(\ell_0,\bar\ell_0)+V_{\rm quark}(\ell_0,\bar\ell_0)\,.\label{eq:V2}
\eeq
The difference with the discussion above is that $\tilde V_{\rm glue}(\bar r_3,\bar r_8)$ is invariant under the transformations (\ref{eq:transfos}) where $r$ has been replaced by $\bar r$. This implies that the definition of $\tilde V_{\rm glue}(\ell_0,\bar\ell_0)$ does not depend on the sector considered for the inversion of (\ref{eq:l0}) and (\ref{eq:lb0}). The rest of the reasoning is the same as above. In particular, the physical Polyakov loops are obtained from functions $\ell(\bar r_3,\bar r_8)$ and $\bar\ell(\bar r_3,\bar r_8)$ that extend $\ell_0(\bar r_3,\bar r_8)$ and $\bar\ell_0(\bar r_3,\bar r_8)$ and which obey a relation similar to (\ref{eq:l0center}) and (\ref{eq:lb0center}) in the heavy quark limit. It follows then that ${\Delta Q\simeq\Delta Q_0}$ obey the same behavior as the one described in the main text as ${T\to 0}$ and ${|\mu|<M}$.

The third and final remark is that the Polyakov loop potential $V(\ell,\bar\ell)$ discussed in the main text is, to some extent, not that different from $V_{\bar r_3,\bar r_8}(\ell_0,\bar\ell_0)$ or $\tilde V(\ell_0,\bar\ell_0)$ defined in Eqs.~(\ref{eq:V1}) and (\ref{eq:V2}) respectively. Indeed, it is shown in Ref.~\cite{Dumitru:2013xna}, see also Re.~\cite{Guo:2018scp}, that $V(\ell,\bar\ell)$ can be extracted from a constrained partition function. The latter, ${W(\bar r_3,\bar r_8)\equiv\ln Z(\bar r_3,\bar r_8)}$, can be computed using the Landau-DeWitt gauge where the backgrounds $\bar r_3$ and $\bar r_8$ are adjusted so that the arguments $\ell$ and $\bar\ell$ of the Polyakov loop potential are identically given by
\beq
\ell(\bar r_3,\bar r_8) & \!\!\!\!=\!\!\!\! & \frac{e^{-i\frac{\bar r_8}{\sqrt{3}}}+2e^{i\frac{\bar r_8}{2\sqrt{3}}}\cos(\bar r_3/2)}{3},\\
\bar\ell(\bar r_3,\bar r_8) & \!\!\!=\!\!\! & \frac{e^{i\frac{\bar r_8}{\sqrt{3}}}+2e^{-i\frac{\bar r_8}{2\sqrt{3}}}\cos(\bar r_3/2)}{3}\,.
\eeq
All in all, one has
\beq
W(\bar r_3,\bar r_8)=V(\ell(\bar r_3,\bar r_8),\bar\ell(\bar r_3,\bar r_8))\,.
\eeq
As before, $W(\bar r_3,\bar r_8)$ is invariant under the transformations (\ref{eq:transfos}) where $r$ has been replaced by $\bar r$. This implies that $V(\ell,\bar\ell)$ can be defined non-ambiguously. In summary, the various approaches presented above differ in the choice of background defining the Landau-deWitt gauge. The background can either be chosen confining, adjusted to equal the gauge-field averages, or adjusted to reproduce the arguments $\ell$ and $\bar\ell$ of the Polyakov loop potential from the above tree-level expressions.

\begin{figure}
    \centering
    \includegraphics[width=.8\linewidth]{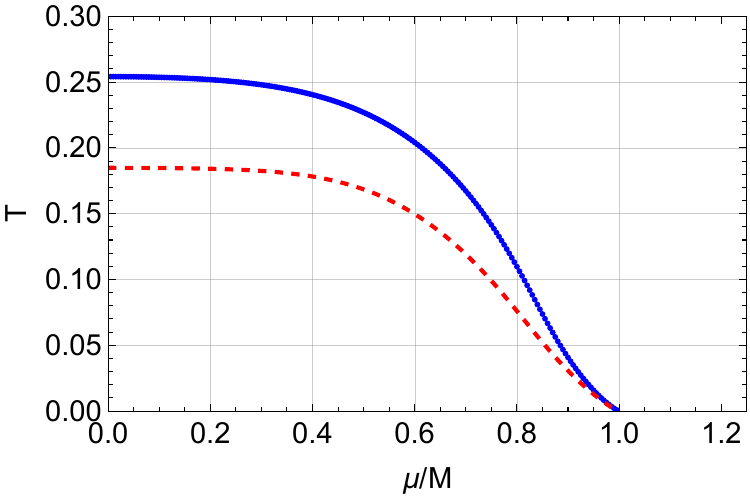}
    \caption{Blue: boundary of the region where $\smash{\ell=\bar\ell=1}$ for the model of Ref.~\cite{Reinosa:2014ooa} where the quark contribution dominates over the glue contribution at low temperatures. Red: confinement/deconfinement transition (with some definition of the crossover transition)}
    \label{fig:llb_self}
\end{figure}

\def\thefigure{A.\arabic{figure}}
\setcounter{figure}{0}

\section{Detailed discussion in the case ${|\mu|>M}$}\label{app:large_mu}

As mentioned in the main text, the discussion of the ${T\to 0}$ limit in the ${|\mu|>M}$ case depends on the strength of the glue contribution to the Polyakov loop potential with respect to the $T^2$-suppressed quark contribution. Let us now analyze each case separately and argue that the final outcome and interpretation for the observables $\Delta Q_q$ and $\Delta Q_{\bar q}$ will be the same in each case.

For models, such as those in Refs.~\cite{Ratti:2005jh,Lo:2013hla}, where the glue contribution dominates,\footnote{For the model of Ref.~\cite{Ratti:2005jh}, the glue potential scales as $T$ for low $T$, whereas it scales as $T^{-1.86}$ for the model of Ref.~\cite{Lo:2013hla}.} we deduce once more that ${(\ell,\bar\ell)\to(0,0)}$. We can then apply the same reasoning as above, and, after linearizing the equations, we arrive this time at
\beq
\ell\sim\bar\ell\simeq\frac{\tilde C}{\partial_\ell\partial_{\bar\ell} V_{\rm glue}}\left|\frac{\mu}{M}\right|\,\left(\frac{\mu^2}{M^2}-1\right)^{1/2},\label{eq:rpml}
\eeq
with ${\tilde C\equiv 2/(\pi\sqrt{3})N_fT^2M^3}$ and where we note that the $T$- and $\mu$-dependencies have factorized. This implies that $\Delta Q_q$ and $\Delta Q_{\bar q}$ both vanish linearly with $T$:
\beq
\Delta Q_q\sim\Delta Q_{\bar q}\simeq \frac{T}{\mu}\frac{2\mu^2-M^2}{\mu^2-M^2}\,.\label{eq:rpml2}
\eeq
We have checked these asymptotic estimates numerically for the models of Refs.~\cite{Ratti:2005jh, Lo:2013hla}.\footnote{It should be stressed that, because ${\ell\sim\bar\ell}$, the linear approximation used to derive Eqs.~(\ref{eq:rpml}) and (\ref{eq:rpml2}) does not suffer from the same problem as in the case ${|\mu|<M}$. We find, however, that the linear approximation is not the most accurate, as one needs to go to rather low temperatures to start seeing the asymptotic estimates (\ref{eq:rpml}) and (\ref{eq:rpml2}). We also mention that, in deriving the simple estimates (\ref{eq:rpml}) and (\ref{eq:rpml2}), we neglected terms of order $T^2\ell$ or $T^2\bar\ell$. We could add them back to the price of obtaining more complicated expressions, but without improving much the quality of the approximation.} Note that, although $\ell$ and $\bar\ell$ approach $0$ as ${T\to 0}$, this occurs at a way smaller (power law) rate than the exponential rate found in the case ${|\mu|<M}$. At low temperatures, the Polyakov loops are then always way smaller in the region ${|\mu|<M}$ than in the region ${|\mu|>M}$, and the latter can be interpreted as the deconfined phase. The vanishing of $\Delta Q_{q/\bar q}$ in this region is then compatible with the interpretation that, in this case, it is possible to bring a color probe without significantly affecting the net quark number of the medium.

In models, such as the one in Ref.~\cite{Reinosa:2014ooa}, where the quark contribution dominates over the glue contribution, it would seem natural to neglect the glue contribution in Eqs.~(\ref{eq:g1}) and (\ref{eq:g2}). The problem when doing so is that the equations ${\partial V_{\rm quark}/\partial\ell=0}$ and ${\partial V_{\rm quark}/\partial\bar\ell=0}$ seem to feature no solution, at least numerically, and thus the solution seems to be lost. Numerical estimates in the model of Ref.~\cite{Reinosa:2014ooa}  indicate that this is not only a problem of the quark contribution but actually of the whole Polyakov loop potential $V(\ell,\bar\ell)$ when ${|\mu|>M}$. In the region ${|\mu|<M}$, while a solution is found, it seems to approach ${(\ell,\bar\ell)=(1,1)}$ as $\mu$ or $T$ increase, until the solution is lost.

We note that ${(\ell,\bar\ell)=(1,1)}$ is a very special point since the corresponding point ${(\bar r_3,\bar r_8)=(0,0)}$ is an extremum of $W(\bar r_3,\bar r_8)$, see the end of the previous section, while ${(\ell,\bar\ell)=(1,1)}$ is not an extremum of $V(\ell,\bar\ell)$. The disappearance of the solution at ${(\ell,\bar\ell)=(1,1)}$ can then be interpreted as resulting from the fusion of the relevant extremum of $W(\bar r_3,\bar r_8)$ with the extremum at ${(\bar r_3,\bar r_8)=(0,0)}$ for large enough $T$ or $\mu$. For larger values of these parameters, the solution gets trapped at ${(\bar r_3,\bar r_8)=(0,0)}$ corresponding to ${(\ell,\bar\ell)=(1,1)}$. In the case of the model of Ref.~\cite{Reinosa:2014ooa}, the discussion should be done in terms of $V_{\bar r_3,\bar r_8}(\bar r_3,\bar r_8)$ rather than $W(\bar r_3,\bar r_8),$ but the interpretation is the same. Fig.~\ref{fig:llb_self} represents, for this model, the boundary of the region of the phase diagram where ${\ell=\bar\ell=1}$. This region contains the entirety of ${|\mu|>M}$. Then, in this region, we have  $\Delta Q_{q,\bar q}$ strictly equal to $0$, which is compatible with what one would expect from a deconfined phase.

\begin{figure}
    \centering
    \includegraphics[width=.8\linewidth]{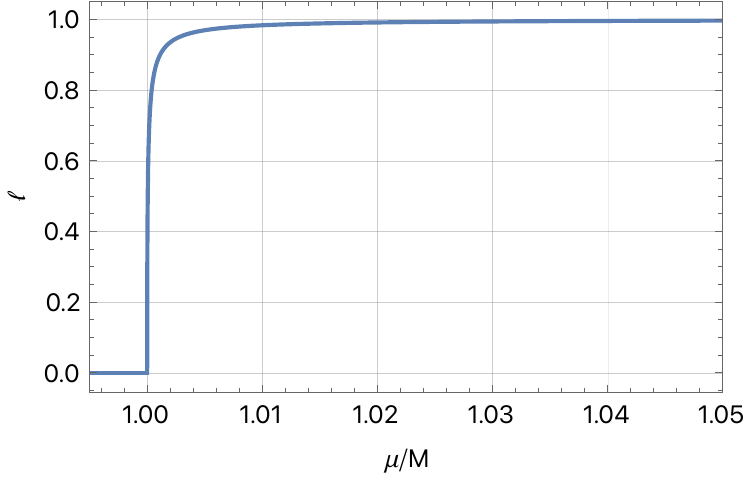}
    \caption{Low-temperature Polyakov loops in the region $\smash{|\mu|>M}$ for the model of Ref.~\cite{MariavanEgmond2022ATemperature} where the quark contribution scales as the glue contribution at low temperatures.}
    \label{fig:llb_cs}
\end{figure}

Finally, in models such as the one in Ref.~\cite{MariavanEgmond2022ATemperature}, where the glue contribution scales as the quark contribution,\footnote{At least, in the absence of renormalization group improvements.} the Polyakov loops converge to a value which is not necessarily equal to $1$ but is determined from the equation
\beq
\frac{\partial v_{\rm glue}(\ell,\ell)}{\partial\ell} & \!\!\!\!\simeq\!\!\!\! & \frac{6N_f}{\pi^2}|\mu|\,(\mu^2-M^2)^{1/2}\,\nonumber\\
& & \times\,\int_{-\infty}^{\infty} \frac{dx\,e^{-x}}{1+e^{-3x}+3\ell(e^{-x}+e^{-2x})},
\eeq
where $V_{\rm glue}(\ell,\bar\ell)\simeq T^2v_{\rm glue}(\ell,\bar\ell)$ and we have assumed that ${\bar\ell=\ell}$ as discussed above. The value of $\ell$ in the limit ${T\to 0}$ can then have some $\mu$-dependence. But, as long as it is non-zero, this implies that $\Delta Q_q$ and $\Delta Q_{\bar q}$ vanish linearly with $T$. As an illustration consider the model of Ref.~\cite{MariavanEgmond2022ATemperature}, for which
\beq
v_{\rm glue}(\ell,\ell)=\frac{m^2}{2g^2}\left(2{\rm Arcos}\,\left(\frac{3\ell-1}{2}\right)-\frac{4\pi}{3}\right)^2\,.
\eeq
For the parameters $m\simeq 510$\,MeV and $g\simeq 5$, and a quark mass $M\simeq 2$\, GeV, we find the result shown in Fig.~\ref{fig:llb_cs}. The Polyakov loop starts at $0$ for $\mu\to M^+$ but very quickly approaches $1$ as $\mu$ increases away from $M$. These two asymptotic values can be easily deduced from the equation. How fast $0$ and $1$ are connected depends on the parameters. For those considered here, the situation is, in a sense, rather similar to the one with the model of Ref.~\cite{Reinosa:2014ooa}. Strictly speaking, the model of Ref.~\cite{MariavanEgmond2022ATemperature} is based on the gauge-field effective potential with confining backgrounds, see the previous section, so the present conclusions concern $\ell_0$ and $\Delta Q_0$. But this implies that $\ell(\ell_0)$ is a function that changes very rapidly from a small value to a constant $\ell(1)$ in the vicinity of ${|\mu|=M^+}$. It follows that ${\Delta Q\simeq 0}$ as ${T\to 0}$ and ${|\mu|>M}$.

\bibliographystyle{elsarticle-num}
\bibliography{refs}
\end{document}